\documentclass[prl,twocolumn,showpacs,nofootinbibs,superscriptaddress
]{revtex4}
\usepackage{epsf}
\usepackage{amsmath}

\def\beq{\begin{equation}}
\def\eeq{\end{equation}}
\def\bea{\begin{eqnarray}}
\def\eea{\end{eqnarray}}
\def\beqa{\begin{equation}\begin{array}{l}}
\def\eeqa{\end{array}\end{equation}}
\def\eqlab#1{\label{eq:#1}}
\def\figlab#1{\label{fig:#1}}

\def\eref#1{(\ref{eq:#1})}
\def\Eqref#1{Eq.~(\ref{eq:#1})}

\def\Figref#1{Fig.~\ref{fig:#1}}

\def\barr{\left(\begin{array}{c}}
\def\earr{\end{array}\right)}
\def\bmat{\left(\begin{array}{cc}}
\def\emat{\end{array}\right)}
\def\al{\alpha}

\def\ga{\gamma} 
 \def\De{\Delta}\def\vDe{\varDelta}

\def\w{\omega}

\def\pa{\partial}

\def\pa{\partial}

\def\nn{\nonumber}

\def\lag{{\mathcal L}}

\def\mathscr{\mathcal}

\def\3d{3-D}

\def\ol#1{\overline{#1}}

\begin{document}
\preprint{WM-06-111}
\preprint{JLAB-THY-06-}

\title{
New large-$N_c$ relations among the nucleon and nucleon-to-$\Delta$ GPDs}

\author{Vladimir Pascalutsa}
\email{vlad@ect.it}
\affiliation{Physics Department, College of William and Mary,
Williamsburg, VA 23187, USA}
\affiliation{Theory Center, Thomas Jefferson National Accelerator Facility, 
Newport News, VA 23606, USA}
\affiliation{ECT*, Villa Tambosi, Strada delle Tabarelle 286,
I-38050 Villazzano (Trento), Italy}

\author{Marc Vanderhaeghen}
\email{marcvdh@jlab.org}
\affiliation{Physics Department, College of William and Mary,
Williamsburg, VA 23187, USA}
\affiliation{Theory Center, Thomas Jefferson National Accelerator Facility, 
Newport News, VA 23606, USA}

\date{\today}

\begin{abstract}
 We establish relations which express 
the generalized parton distributions (GPDs) describing the 
$N\to \Delta$ transition in terms of the nucleon GPDs. 
These relations are based on the known large-$N_c$ relation
between the $N\to \De$ electric quadrupole moment and the
neutron charge radius, and a newly derived 
large-$N_c$ relation between the electric quadrupole ($E2$) and
Coulomb quadrupole ($C2$) transitions. Namely, in the large-$N_c$
limit we find $C2=E2$.
The resulting relations among the nucleon and $N\to\Delta$ GPDs 
provide predictions for the $N\to\Delta$
electromagnetic form factors which are found to be in very good agreement
with experiment for moderate momentum transfers.
\end{abstract}


\maketitle
\thispagestyle{empty}

Electromagnetic form factors (FFs) are the standard source of information
on the structure of the nucleon and as such have been studied extensively.
The phenomenon of asymptotic freedom in QCD, however, 
provides us with more sophisticated tools of describing the quark-gluon structure of hadrons. 
Generalized parton distributions (GPDs) provide the distribution of quarks
as a function of both their momentum fraction and transverse position in the 
nucleon, see Refs.~\cite{Ji:1998pc,Goeke:2001tz,Diehl:2003ny,Belitsky:2005qn} 
for reviews. 

GPDs can be accessed by selecting a small size configuration of quarks and gluons, 
provided by hard exclusive reactions 
such as deeply virtual Compton scattering (DVCS).  
Figure~\ref{fig:ndelta_dvcs} 
illustrates the leading ("handbag") contribution to DVCS processes with the nucleon or the nucleon
first excitation --- $\Delta$(1232) --- in the final state.
The crucial feature of such hard reactions is the possibility
to separate the perturbative and nonperturbative stages of
the interactions due to the {\it factorization theorems}~\cite{Ji98a,Col99,Rad98}. 
The non-perturbative stage of these hard exclusive processes is described by GPDs. 
 First DVCS experiments aimed to measure GPDs have recently 
been completed, many others are underway.
\begin{figure}
\centerline{  \epsfxsize=6cm
  \epsffile{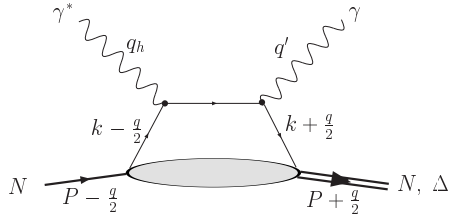} 
}
\caption{
The ``handbag'' diagram for the $N \to N$ and $N \to \Delta$ DVCS processes. 
The virtual photon momentum $q_h$ is the hard scale.
Factorization theorem allows to present this process as the convolution  
of a Compton scattering at the quark level and a non-perturbative 
amplitude (denoted here by the blob) parameterized in terms of GPDs.    
}
\figlab{ndelta_dvcs}
\end{figure}

In this Letter we would like to establish general relations, based on the 
large number-of-colors ($N_c$) limit of QCD, 
between the $N \to N$ and $N \to \Delta$ FFs and GPDs.
The nucleon GPDs are already well-constrained by the knowledge of the
nucleon elastic FFs and 
forward parton distributions. Viable parametrizations
of the nucleon GPDs exist in the literature~\cite{Diehl:2004cx,guidal}. The 
GPDs describing the $N \to \Delta$ transition are, on the other hand, 
poorly known.  
While a comprehensive analysis of the $N \to \Delta$ FFs
exists (see \cite{PVY06} for review), the
corresponding forward parton disributions cannot be measured by means
of deep inelastic
scattering. 

Recently, the possibility of accessing the $N \to \Delta$
GPDs in DVCS was proposed, see~\cite{GMV03} for detailed estimates, and a first
measurement of such a process has already been reported~\cite{Guidal:2003ji}.
Furthermore, by generalizing the large-$N_c$ relations between the
octet and decuplet baryons, it bacame possible to relate the dominant
magnetic dipole ($M1$) $N\to \De$ transition GPD to the nucleon isovector 
GPD~\cite{Frankfurt:1999xe}. In this work we 
look for the large-$N_c$ relations for the electric ($E2$) and Coulomb ($C2$)
quadrupole $N\to \De$ GPDs.  

To introduce the electromagnetic $N\to \De$ transition it is useful
to write down the effective Lagrangian:
\bea
\eqlab{lagran}
\lag_{\ga N\De} &=&   \frac{3 i e}{2M_N (M_N + M_\Delta)} \,\ol N \, T^3 \nn\\
&\times & \left[  g_M \,\pa_{\mu}\De_\nu\, \tilde F^{\mu\nu}  
 + i g_E \,\ga_5\,\pa_{\mu}\De_\nu \, F^{\mu\nu} \right. \\
&- & \left.  \,\frac{g_C}{M_\De} \ga_5 \ga^\al  
(\pa_{\al}\De_\nu-\pa_\nu\De_\al) \,\pa_\mu F^{\mu\nu}\right] + \mbox{H.c.},
\;\;\;\;\; \nn
\eea
where $N$ denotes the nucleon (spinor) 
and $\De_\mu$  the $\De$-isobar (vector-spinor) fields, 
$M_N$ and $M_\De$ are respectively their masses, $F^{\mu\nu}$ and $\tilde F^{\mu\nu}$
are the electromagnetic field strength and its dual, 
$T^3$ is the isospin-1/2-to-3/2 transition operator. 
The important observation that we take
from this expression is that the couplings $g_M$, $g_E$ and $g_C$ appear with the same
structure of spin-isospin and field operators, and hence should scale with the same
power of $N_c$, for large $N_c$.

It is customary to characterize the three different 
types of the $\ga N \De$ transition
in terms of the Jones--Scadron FFs $G^*_M$,  $G^*_E$,
$G^*_C$. The contribution of the effective couplings \Eqref{lagran} 
to these FFs can be
straightforwardly computed with the following result:
\begin{eqnarray}
\label{eq:JS}
G_M^\ast(Q^2) &=& g_M \,+\left[ -M_\De \,\w\,g_E
+  Q^2 g_C\right]/Q_+^2\,\,, \nn\\ 
G_E^\ast(Q^2) &=& \left[-M_\De \,\w\, g_E+  Q^2  g_C\right]/Q_+^2\,\,, \\
G_C^\ast(Q^2) &=&-2M_\De\, \left[ \w \,  g_C + M_\De \,g_E\right]/Q_+^2\,\,,\nn
\end{eqnarray}
where we introduced $Q_\pm = \sqrt{(M_\De\pm M_N)^2 +Q^2}$ and the 
photon energy in the $\Delta$ rest frame: $\w=(M_\De^2-M_N^2-Q^2)/(2M_\De)$. 
We immediately note that at $Q^2 = 0$, 
\begin{subequations}
\bea
\eqlab{GEstar}
G_E^\ast(0) &=& - \mbox{$\frac{\vDe}{2 (M_N+M_\De)}$}  \, g_E , \\
G_C^\ast(0) &=&-\mbox{$\frac{2M_\De^2}{(M_N+M_\De)^2}$} 
\left[ \mbox{$\frac{M_N+M_\De}{2M_\De}\,\frac{\vDe}{M_\De}$} \,  g_C +  g_E\right],
\eqlab{GCstar}
\eea 
\end{subequations}
where $\vDe \equiv M_\De - M_N$ is the $\De$-nucleon mass difference. 
In the large-$N_c$ limit this mass difference goes as $1/N_c$, 
whereas the baryon masses increase proportionally to $N_c$:
\beq
M_{N (\De)} = {\mathcal O}(N_c),\,\,\, \vDe = {\mathcal O}(N_c^{-1}) \,.
\eeq
Given the fact that $g_E$ and $g_C$ scale with the same power of $N_c$, we observe
that the first term in \Eqref{GCstar} is suppressed by $1/N_c^2$ and therefore obtain
the following large-$N_c$ relation:
\beq
G_C^\ast(0) \,=\, \frac{2 M_\Delta}{M_N+M_\Delta} \,\frac{2M_\De}{\vDe} \,G_E^\ast(0)\,.
\label{eq:gc0largenc}
\eeq

Of special interest are the multipole ratios: $R_{EM}=E2/M1$ and $R_{SM}=C2/M1$,
which can be expressed in terms of the Jones-Scadron FFs:
\beq
\eqlab{ratios}
R_{EM} = - \frac{G_E^\ast}{G_M^\ast} \,,\quad \quad 
R_{SM} = - \frac{Q_+ Q_-}{4M_\De^2} \, \frac{G_C^\ast}{G_M^\ast}.
\eeq
It is easy to see that our relation \eref{gc0largenc} translates into
$R_{SM} = R_{EM}$,  at $Q^2=0$. Using Eqs.~\eref{GEstar} and \eref{ratios}
one readily verifies the result of Ref.~\cite{Jenkins:2002rj}: $R_{EM}={\mathcal O}(1/N_c^2)$.

We now turn to GPDs and recall that in general they depend on
three variables: $x$, $\xi$, and $Q^2$.
The light-cone momentum fraction $x$ is defined by $k^+ = x P^+$,
where $k$ is the quark loop momentum and
$P$ is the average nucleon momentum 
$P = (p + p^{\ \prime})/2$, where $p (p^{\ \prime})$
are the initial (final) baryon four-momenta,   
see \Figref{ndelta_dvcs}.
The skewedness variable $\xi$ is
defined by $q^+ = - 2 \xi \,P^+$, where 
$q = p^{\ \prime} - p$ is the
overall momentum transfer in the process, and where
$2 \xi \rightarrow x_B/(1 - x_B/2)$ in the Bjorken limit;  
$x_B = Q_h^2/(2 p \cdot q_h)$ is the usual Bjorken scaling variable, 
with $Q_h^2 = -q_h^2 > 0$ the virtuality of the hard photon.
Finally, the third variable
is the total momentum transfer squared: $Q^2 = - q^2$.
 
In a frame where the virtual photon momentum \( q_h \) and the average
nucleon momentum \(  P \) are along the \( z \)-axis and 
opposite to each other, one can parameterize
the non-perturbative piece of the $N \to \Delta$ DVCS amplitude 
as~\cite{Goeke:2001tz,Frankfurt:1999xe}:
\begin{eqnarray}
&& \frac{1}{2\pi} \int dy^{-}e^{ix  P^{+}y^{-}}
\langle \Delta |\bar{\psi } (-\frac{y}{2}) \, \gamma \cdot n  \, \tau_3  
\psi (\frac{y}{2})
 | N \rangle {\Bigg |}_{y^{+}=\vec{y}_{\perp }=0} \nonumber \\
&& =  \sqrt{2/3} \; u^{\alpha }(p^\prime)\; 
\left\{ \, H_{M}(x,\xi ,Q^2)\; 
\left( -{\mathcal{K}}_{\alpha \mu }^{M} \right)  \right. 
\nonumber \\
&&\hspace{2.cm}  +\; H_{E}(x,\xi ,Q^2)\; 
\left( -{\mathcal{K}}_{\alpha \mu }^{E} \right) \nonumber \\
&&\hspace{2.cm}  +\; H_{C}(x,\xi ,Q^2)\; 
\left( -{\mathcal{K}}_{\alpha \mu }^{C} \right) \nonumber \\
&&\hspace{2.cm} \left. +\; H_{4}(x,\xi ,Q^2)\; 
\left( g_{\alpha \mu } \, \gamma_5 \right) \, \right\} \, n^{\mu} \, u(p), 
\label{eq:ndelvec} 
\end{eqnarray}
where \( \psi  \) is the quark field
of flavor $q$, \( u \) the nucleon spinor, 
\( u ^{\alpha }(p^\prime) \) is the Rarita-Schwinger spinor
for the \( \Delta  \)-field, 
$n^\mu$ is a light-cone vector along the negative $z$-direction, 
$\tau_3 /2$ is the third isospin generator for quarks, 
and $\sqrt{2/3}$ is the isospin
factor for the $p \to \Delta^+$ transition. 
Furthermore,
the covariants \( {\mathcal{K}}^{M,E,C}_{\alpha \mu } \)
are~\cite{Jones:1972ky}:
\begin{eqnarray}
{\mathcal{K}}_{\alpha \mu }^{M} & = & 
-i\frac{3(M_{\Delta }+M_N)}{2M_N Q_+^2} 
\varepsilon_{\alpha \mu \lambda \sigma} 
P^{\lambda } q^{\sigma}\; ,\nonumber \\
{\mathcal{K}}_{\alpha \mu }^{E} & = & -{\mathcal{K}}_{\alpha \mu }^{M}-\frac{6(M_{\Delta }+M_N)}{M_N Q_+^2 Q_-^2 }
\varepsilon _{\alpha \sigma \lambda \rho }P^{\lambda } q^{\rho } \, 
\varepsilon^{\sigma}_{\, \, \mu \kappa \delta }P^{\kappa} q^{\delta}\gamma_5\;,
\nonumber \\
{\mathcal{K}}_{\alpha \mu }^{C} & = & 
-i\frac{3(M_{\Delta}+M_N)}{M_N Q_+^2 Q_-^2} 
q_{\alpha }(q^2 P_{\mu }- q \cdot P q_{\mu })\gamma_5 \; .
\label{K-def} 
\end{eqnarray}
The thus introduced GPDs \( H_{M} \), \( H_{E} \),
and \( H_{C} \) correspond with the 
three \( N\rightarrow \Delta  \) Jones--Scadron FFs
and relate to them via the sum rules: 
\begin{eqnarray}
\int _{-1}^{1}dx\, \, H_{M,E,C}(x,\xi ,Q^2)
=2 \, G_{M,E,C}^{*}(Q^2)\; .
\label{eq:gmgegcsumrule} 
\end{eqnarray}
The fourth $N \to \Delta$ vector GPD $H_4$ in 
\Eqref{ndelvec} has a vanishing first moment.  
\newline
\indent 
For the magnetic dipole $N \to \Delta$ transition, 
it was shown that, in the large-$N_c$ limit, the 
relevant  $N \to \Delta$ GPD $H_M$ can be expressed in terms of the
nucleon isovector GPD $E^u - E^d$ as~\cite{Frankfurt:1999xe}~: 
\begin{eqnarray}
H_{M} & = & 2  \frac{G_M^*(0)}{\kappa_V}  
 \left\{ E^{u} - E^{d} \right\} ,
\label{eq:hmparam} 
\end{eqnarray}
where $\kappa_V = \kappa_p - \kappa_n = 3.70$ is the nucleon isovector 
anomalous magnetic moment. 
The large-$N_c$ limit value 
$G_M^{*}(0) = \kappa_V / \sqrt{2}$ is about  20\% smaller than the
experimental number, and therefore we will use 
the phenomenological value $G_M^{*}(0) \approx 3.02$~\cite{Tiator:2003xr} 
in the calculations below.  
\newline
\indent
Using the large-$N_c$ estimate of Eq.~(\ref{eq:hmparam}), the sum rule 
Eq.~(\ref{eq:gmgegcsumrule}) for $G_M^\ast$ can be written as:    
\begin{eqnarray}
G_M^{*}(Q^2) &=& {{G_M^{*}(0)} \over {\kappa_V}} \int _{-1}^{+1}dx 
\left[ E^{u} - E^{d} \right] (x,\xi ,Q^2) , 
\nonumber \\
&=&  {{G_M^{*}(0)} \over {\kappa_V}} \; 
\left\{ F_2^p(Q^2) - F_2^n(Q^2) \right\} \, ,  
\label{eq:gmsumrule} 
\end{eqnarray}
where $F_2^p - F_2^n$ is the (isovector) combination of the 
proton (p) - neutron (n) Pauli FFs. 
Because the sum rule~(\ref{eq:gmsumrule}) is independent of $\xi$, 
we only need to constrain the GPD $E^q$ for $\xi = 0$ in order to 
evaluate $G_M^\ast$.  
\begin{figure}
\centerline{  \epsfxsize=9cm%
  \epsffile{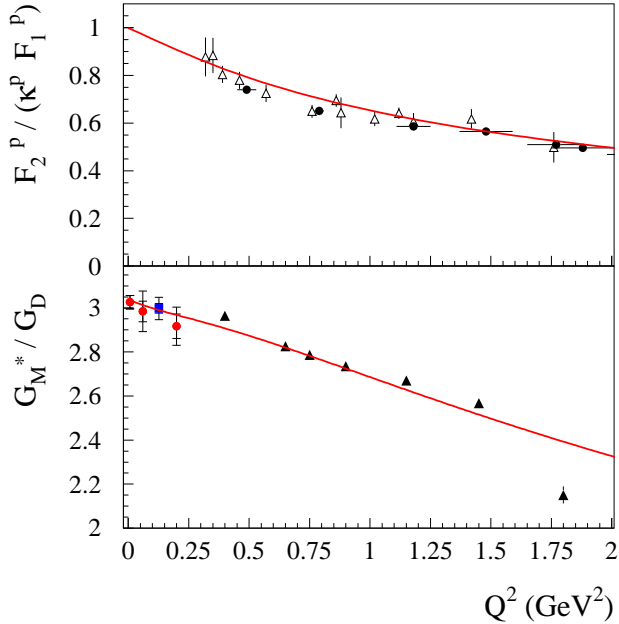} 
}
\caption{(Color online)
Upper panel: proton FF ratio $F_2^p / (\kappa^p F_1^p)$. 
Solid curve --- ``modified Regge'' parametrization~\cite{guidal}. 
Data points are from JLab/Hall A:~\cite{Gayou:2001qt} (open triangles), 
and \cite{Punjabi:2005wq} (solid circles). 
Lower panel: $N \to \Delta$ (Jones-Scadron) 
form factor $G_M^*$, relative to the dipole form $G_D = 1/(1 + Q^2/0.71)^2$. 
The curve is calculated from the large-$N_c$ prediction 
(\ref{eq:gmsumrule}) for the $N \to \Delta$ GPD $H_M$. 
Data points are from 
MAMI~\cite{Beck:1999ge,Stave:2006ea,Sparveris:2006} (solid circles),  
MIT-Bates~\cite{Sparveris:2004jn} (solid square), 
and JLab/CLAS~\cite{Joo:2001tw} (solid triangles).}
\label{fig:f2gmdel}
\end{figure}
In Ref.~\cite{guidal}, a 3-parameter modified Regge parametrization 
for the nucleon GPDs was found to provide a good quantitative 
description for all four nucleon elastic form factors over the whole $Q^2$ 
range. This is illustrated in Fig.~\ref{fig:f2gmdel} 
for the ratio of the proton Pauli over Dirac FFs. 
Using the sum rule prediction based on the large $N_c$ estimate of 
Eq.~(\ref{eq:gmsumrule}), the $N \to \Delta$ magnetic FF $G_M^\ast$ follows 
as a prediction. It is seen from Fig.~\ref{fig:f2gmdel} that the large $N_c$ 
relation reproduces well the 
experimentally observed, faster than dipole, fall-off of $G_M^\ast$.  
One sees that this is also in agreement with the corresponding 
fall-off of the nucleon isovector Pauli form factor, 
confirming the finding of Ref.~\cite{Stoler:2002im} using a Gaussian model 
for the GPDs. 
\newline
\indent
We shall now relate the $N \to \Delta$ GPDs for the electric 
quadrupole ($H_E$) 
and Coulomb quadrupole ($H_C$) transitions with the nucleon GPDs. 
We start from a large $N_c$ relation between the $N \to \Delta$ 
quadrupole moment $Q_{p \to \Delta^+}$ 
and the neutron charge radius $r_n^2$ \cite{Buchmann:2002mm}:
\begin{eqnarray} 
Q_{p \to \Delta^+} \,=\, \frac{1}{\sqrt{2}} \, r_n^2 .
\label{eq:qpdelrnlargenc2}
\end{eqnarray}
Using the Jones--Scadron $N \to \Delta$ form factors, 
we can express Eq.~(\ref{eq:qpdelrnlargenc2}) 
as a relation for $G_E^*(0)$, which reads (to leading order in the 
$1/N_c$-expansion) as:
\begin{eqnarray}
G_E^\ast(0) \,=\, - \,
\frac{M_\Delta^2 - M_N^2}{12\sqrt{2}}\,r_n^2 .
\label{eq:ge0largenc}
\end{eqnarray} 
Note that the experimental value of the $R_{EM}$ ratio at the real photon point 
($R_{EM} = - 2.5 \pm 0.5 \%$ \cite{PDG2006}) yields~:
$Q_{p \to \Delta^+} = -( 0.085 \pm 0.003 )$~fm$^2$~\cite{Tiator:2003xr}.   
Using the experimental neutron charge radius $r_n^2 = -0.113$~fm$^2$, 
the large-$N_c$ relation of Eq.~(\ref{eq:qpdelrnlargenc2}) 
yields: $Q_{p \to \Delta^+}$ = $-0.08$~fm$^2$, in close agreement with
experiment. 

For small values of $Q^2$, the neutron electric FF is expressed   
as $G_E^n(Q^2) \approx - r_n^2 \, Q^2 / 6$. Therefore, an extension 
of the large-$N_c$ relation (\ref{eq:ge0largenc}) to finite $Q^2$ 
is given by:
\begin{eqnarray}
G_E^\ast(Q^2) \simeq 
\frac{1}{\sqrt{2}} \, \frac{M_\Delta^2 - M_N^2}{2Q^2} \, 
G_E^n(Q^2) .
\label{eq:geqlargenc}
\end{eqnarray}
\newline
\indent
An analogous prediction for the Coulomb quadrupole 
$N \to \Delta$ GPD, $H_C$, can be made by using the 
newly derived relation~\Eqref{gc0largenc}.
Extending this relation to finite $Q^2$, we have
\begin{eqnarray}
G_C^\ast (Q^2) \simeq \frac{4 M_\Delta^2}{M_\Delta^2 - M_N^2} \, 
G_E^\ast(Q^2). 
\label{eq:gcqlargenc}
\end{eqnarray} 
Using \Eqref{ratios}, we can turn this into a relation between the 
$R_{EM}$ and $R_{SM}$ ratios as~:
\begin{eqnarray}
R_{SM}(Q^2) \,\approx\, \frac{Q_+ \, Q_-}{M_\Delta^2 - M_N^2} \, 
R_{EM}(Q^2) .
\label{eq:rsmlargenc}
\end{eqnarray}
At $Q^2 = 0$, one recovers the relation 
$R_{SM} = R_{EM}$. 
\newline
\indent
The FFs $G_E^\ast$ and $G_C^\ast$ are 
obtained from the first moment of 
the $N \to \Delta$ GPDs $H_E$ and $H_C$ through 
the sum rules of Eq.~(\ref{eq:gmgegcsumrule}). We can therefore use 
Eqs.~(\ref{eq:geqlargenc}) and~(\ref{eq:gcqlargenc}) to obtain 
relations between the $N \to \Delta$ GPDs $H_E$ and $H_C$ and 
the neutron electric GPD combination as~:
\begin{eqnarray}
H_E \,&=&\, \frac{1}{\sqrt{2}} \, 
\frac{M_\Delta^2 - M_N^2}{Q^2} \, 
\left\{ H^{(n)} - \frac{Q^2}{4 M_N^2} E^{(n)} \right\} , 
\label{eq:helargenc} \nn \\
H_C \,&=&\, \frac{4 M_\Delta^2}{M_\Delta^2 - M_N^2} \, H_E,
\label{eq:hclargenc}
\end{eqnarray}
where the (neutron) GPDs $H^{(n)}$ and $E^{(n)}$ 
are defined in terms of the $u$- and $d$-quark flavor GPDs as: 
$H^{(n)} \equiv -1/3 H^u + 2/3 H^d$, and $E^{(n)} \equiv -1/3 E^u + 2/3 E^d$.

\begin{figure}
\centerline{  \epsfxsize=9cm%
  \epsffile{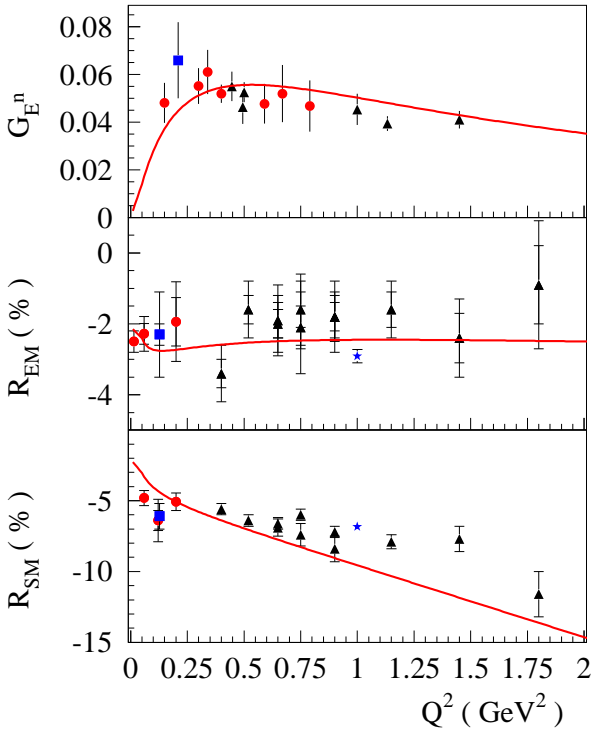} 
}
\caption{(Color online)  
Neutron electric FF $G_E^n$ (upper panel) in comparison with  
the $N \to \Delta$ $R_{EM}$ (middle panel) 
and $R_{SM}$ (lower panel) ratios. The curves result
from the ``modified Regge'' GPD parametrization~\cite{guidal}, where  
in computing $R_{EM}$ and $R_{SM}$, the large-$N_c$ relations 
(\ref{eq:geqlargenc}) for $G_E^\ast$, 
(\ref{eq:gcqlargenc}) for $G_C^\ast$, 
and (\ref{eq:gmsumrule}) for $G_M^\ast$ are used.  
Data points for $G_E^n$ are from 
MAMI~(red circles), 
NIKHEF~(blue square), 
and JLab~(black triangles), see \cite{guidal} for references. 
Data points for $R_{EM}$ and $R_{SM}$ 
are from BATES~\cite{Sparveris:2004jn} (blue squares),   
MAMI~\cite{Beck:1999ge,Stave:2006ea,Sparveris:2006} (red circles), 
JLab/CLAS~\cite{Joo:2001tw} (black triangles),  
and JLab/HallA~\cite{Kelly05} (blue stars). 
}
\figlab{fig:remrsm_gpd}
\end{figure}

The prediction which follows from the large-$N_c$ motivated 
expression of Eq.~(\ref{eq:geqlargenc}) is 
tested in \Figref{fig:remrsm_gpd} by comparing the 
$Q^2$ dependence of the neutron electric FF $G_E^n$ 
and the $N \to \Delta$ $R_{EM}$ and $R_{SM}$ ratios. 
Although the above relations are derived assuming small $Q^2$, 
we explore their empirical validity at moderate $Q^2$ as well. 
The neutron FF $G_E^n$ is computed using the modified Regge parametrization of 
\cite{guidal}, which is seen to give a fairly good description of the 
available double polarization data. The $R_{EM}$ and $R_{SM}$ ratios 
are then computed using 
the large-$N_c$ relations (\ref{eq:geqlargenc}) for $G_E^\ast$, (\ref{eq:gcqlargenc}) for 
$G_C^\ast$, and (\ref{eq:gmsumrule}) for $G_M^\ast$. 

By using the three 
parameter Regge form for the nucleon GPDs, 
we thus have obtained parameter-free predictions for $R_{EM}$ and $R_{SM}$. 
One sees that this yields a $R_{EM}$ ratio 
which has both the right size and displays 
a relatively flat $Q^2$ behavior, up to $Q^2$ of about  
2~GeV$^2$. The prediction of the $Q^2$ dependence for both 
ratios is in a surprisingly (for a large-$N_c$ estimate) 
good agreement with the experimental data. 
\newline
\indent
Summarizing, we derived new large-$N_c$ relations expressing the 
$N \to \Delta$ GPDs in terms of the nucleon GPDs. 
In particular, the $R_{EM}$ and $R_{SM}$ ratios were  
related to the neutron electric form factor. Using a parameterization for the 
nucleon GPDs, these relations were found to yield a surprisingly good 
agreement with experiment, 
providing an explanation why $R_{EM}$ remains nearly constant for $Q^2$ values 
up to several GeV$^2$. 
\newline
\indent
A worthwhile topic for future work is 
to perform model calculations for 
the $N \to \Delta$ GPDs, as well as provide lattice QCD estimates for 
its moments, in order to cross-check the above relations. 

\begin{acknowledgments} 
This work is supported in part by DOE grant no.\
DE-FG02-04ER41302 and contract DE-AC05-06OR23177 under
which Jefferson Science Associates operates the Jefferson Laboratory. 
\end{acknowledgments}

\end{document}